# Solar Type U Burst Associated with a High Coronal Loop

V. V. Dorovskyy[1], V. N. Melnik[1], A. A. Konovalenko[1], S. N. Yerin[1], I. N. Bubnov[1]

[1] – Institute of Radio Astronomy of National Academy of Sciences of Ukraine.

Submitted to Solar Physics 6 August 2020. Accepted 22 November 2020.

## Abstract

An inverted U burst with equally developed ascending and descending branches observed by Giant Ukrainian Radio Telescope (GURT) on 18 April 2017 in meter wavelengths band is discussed. This U burst was attributed to the high coronal loop above the limb active region NOAA 12651. Under the assumption that, associated with the burst, the coronal loop confines isothermal plasma stratified according to a Boltzmann density relation, the geometrical and physical parameters of the loop were estimated. According to our model coronal loops may contain plasma which is up to 20 times denser than the surrounding coronal plasma. In general the proposed model gives the relation between the plasma temperature and the height of the loop in such a way that under the given parameters of the associated U burst, higher loops contain cooler plasma and vice versa. An alternative method of coronal loop height determination was suggested. Assuming that the observed U burst and the preceding Type III burst were generated by the same exciter we define the height of the loop from the delay of the former with respect to the latter at certain frequency. We show that defining of coronal loops heights by another independent method, e.g. interferometric or tied-array imaging may solve the uncertainty of the inside-the-loop plasma temperature determination.

Keywords Radio bursts: dynamic spectrum · Meter-wavelengths and longer (m, dkm, hm, km) · Type III · Corona: structures

## Introduction

Since the very discovery in 1958 (Maxwell and Swarup, 1958) solar inverted U bursts have been considered as a special case of the Type III bursts, which were firstly observed 8 years earlier (Wild, 1950). This time gap seems natural because type U bursts are much more rarely observed compared to type IIIs (Suzuki and Dulk, 1985). Bursts of these two types are cognate primarily due to common exciters of the emission, which are sub-relativistic electron beams accelerated above active regions and traveling along solar magnetic field lines (Suzuki and Dulk, 1985). The decisive difference between them is the shape of guiding magnetic field lines and thus the trajectories of the exciter movement. Type III electrons move along open field lines at distances up to 1 AU generating bursts with monotonically decreasing drift rates defined as type III bursts, which can be continuously traced in frequency from GHz down to kHz. In contrast electrons responsible for inverted U burst generation follow closed magnetic structures – coronal loops. In this case the radial velocity of the electron beam alters its sign to the opposite while passing the apex of a coronal loop causing change of frequency drift sign of the burst from negative to positive and forming the ascending and descending branches of a burst. Thus unlike type III bursts the dynamic spectra of inverted U bursts are bounded by the turning frequencies. These turning frequencies are observed in rather wide frequency range from 3.8GHz (Wang et al., 2001) down to 0.7 MHz (Stone and Fainberg, 1971). Nevertheless the bursts of this type are most common for metric and decametric bands (Suzuki and Dulk, 1985; Dorovskyy, et al., 2010; Reid and Kontar, 2017). More detailed and extended analysis made by Leblanc and Hoyos (1985) showed that U bursts turning frequencies most often occur in frequency range 25-30 MHz.

Ascending branches of U bursts far away from the turning point morphologically resemble typical type III bursts. Besides the fact that they use to drift slightly slower than type III bursts do at the same frequency. Such drift rates were usually attributed to slightly slower exciters (Fokker, 1970; Dorovskyy et al., 2010). When U bursts are observed their descending legs are usually weaker, more diffuse and less extended in frequency than the ascending ones (Poquerusse, Bougeret, and Caroubalos, 1984). In most cases descending legs of U bursts are absent at all resulting in so called J burst. U and J bursts are characteristic of high solar activity periods and sometimes may gather in groups. The turning frequencies of the bursts in one group may remain stable or decrease with time pointing out that some coronal loops are stationary while the others rise with time (Leblanc, Poquerusse, and Aubier, 1983; Dorovskyy et al. 2010).

As well as type III bursts solar U bursts sometimes show harmonic structures consisting of fundamental and harmonic emission with the former component delayed with respect to the latter by a few seconds (Dorovskyy, et al. 2015). Then if single individual J or U bursts are observed it is natural to conclude that they can be generated either at fundamental or at harmonic frequency. Comparative rareness of U bursts with respect to type III bursts is unexpected, given the abundance of loop structures in the solar corona. Reid and Kontar (2017) explained this effect in such a way that electron beams need to path a certain distance to become unstable to the Langmuir waves, which is in most cases larger than linear length of loops. The authors also add that higher density and lower density gradient of the inside-the-loop plasma in turn reduce the Langmuir wave's growth rate thus increasing the starting heights of the U bursts. From this analysis it follows that the emission of U bursts begins far away from the electron injection region and hence observed U bursts apparently correspond to the upper part of the whole coronal loops.

Retrieving of coronal loop geometry from associated U burst parameters has been one of the most relevant tasks in study of this type of bursts. According to Reale (2014) coronal loops can be visible in EUV band if the density of the confined plasma exceeds $10^8$ $cm^{-3}$. Such densities are characteristic of the lower layers of the corona. Thus loops being as high as a few solar radii and containing less dense plasma can be nowadays diagnosed exclusively by the means of radio astronomy. In the absence of spatial parameters measurements the heights of the loops are usually defined by the turn over frequency of associated U burst assuming one of known density profile models (e.g. Stone and Fainberg, 1971; Leblanc and Hoyos, 1985; Dorovskyy, et al. 2010). In most cases these models were obtained for solar corona in general rather than for coronal structures, such as loops or streamers. At the same time spatial measurements performed at different times show that the sources of U bursts are located higher in corona than expected from the models (Suzuki, 1978; Reid and Kontar, 2017) pointing out that density stratification inside these structures most probably differ from that outside them.

The present paper is an attempt to estimate the main parameters of the top part of the coronal loop using the time and frequency properties of the associated U burst with both legs unprecedentedly well developed. The assumptions adopted here are similar to those in the analysis of the U bursts harmonic pair (Dorovskyy et al., 2015).

**Observations and the inverted U burst properties.**

On 18 April 2017 two active regions NOAA 12651 appeared from behind the eastern limb of the Sun. During that day two comparatively powerful flares happened above this region: C2 flare at around 9:30 UT and C5 flare with peak intensity at 20:00 UT. Both of these flares initiated CMEs which travelled eastward according to SOHO/LASCO white light images. Besides the CMEs different types of solar activity in radio band were associated with these flares.

Since solar observations with radio telescope GURT are usually held between 5 and 16 UT we will focus on the events associated with the first C2 flare.

In meter and decameter bands this flare manifested itself by a complex long-lasting event started at 9:34 UT with a type III burst with decay followed by inverted type U burst, group of powerful type III bursts, type II burst with herring-bone structure, and type IV burst observed with Nançay Decameter Array (NDA; Lecacheux, 2013), URAN-2 radio telescope (Brazhenko, et al., 2005) and with one section of newly built Giant Ukrainian Radio Telescope (GURT, Konovalenko, et al., 2016). One section of radio telescope GURT operates in frequency band 10 – 80 MHz and consists of 25 cross-dipoles making up an effective area of 350 $m^2$ and beam width of about 20º at 40MHz. The time and frequency resolutions were set to 100ms and 9 kHz respectively. All these bursts were briefly described in (Melnik et al, 2019). And the type III burst with decay was the subject of the research in that paper. We should only add that the complex event was preceded by a series of inverted J bursts with turning frequencies of around 30 MHz. The series totaled up to 10 individual J bursts and lasted from 9:29:00 till 9:33:30 UT. This fact indicated the existence of rather high coronal loops above the active region as well as the conditions conductive to escaping of radio emission from loop structures. The apexes of these J bursts were also observed by URAN-2 radio telescope providing also the polarization data. All mentioned bursts except type IV burst were weakly polarized with a degree not exceeding 5%, indicating that all type J and type III bursts were most probably generated at the second harmonic of local plasma frequency despite no signs of the fundamental components were visible. The latter fact may happen due to large solar latitude and much narrower radiation pattern of the fundamental emission source.

The fragment of the whole event recorded by Nançay Decameter Array (NDA) is shown in Figure 1. And Figure 2a represents the inverted U burst with equally developed branches observed with GURT. It is clear that due to insufficient sensitivity and low time resolution (1s) of NDA data compared to GURT it does not allow detailed analysis of the bursts of interest. The discussed U burst first appears at 9:34:30 UT at frequency around 75 MHz. The intensity of the observed U burst peaks at low frequency around 55 MHz reaching almost 20 dB above the background level and starts to fall above 70 MHz equalizing with the background level at 76 MHz. This fall is an aggregate effect of the input low-pass filter and antenna efficiency deterioration at high frequencies but not the actual flux fall. The GURT data is still not calibrated so we do not bring the values of fluxes here. Nevertheless the exact values of the burst fluxes are not as important for the present research as frequency and temporal parameters. Equally developed branches of the burst allow to determine the drift rates of the ascending and descending ones at the same frequency rather far away from the turn over point. This unique opportunity is of great importance for retrieving the coronal loop parameters.

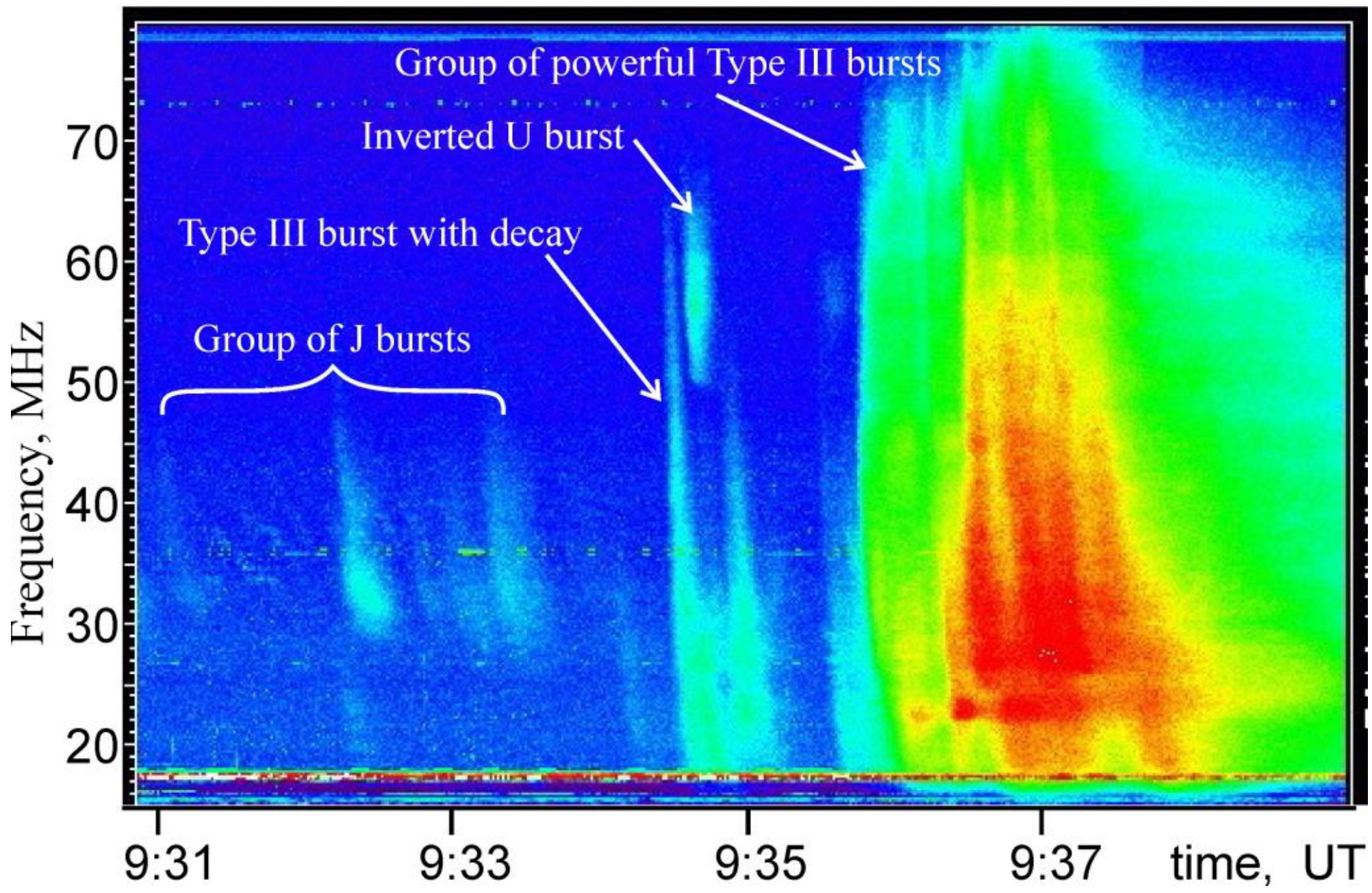

Figure 1. The event associated with the C2 flare on 18 April 2017 recorded by NDA.

Due to low intensity of the burst at high frequencies and overall smooth appearance of the burst accurate measurement of the drift rates and time intervals from power dynamic spectrum (Figure 2a) seems difficult. The differential in time dynamic spectrum (Figure 2b) is better suited for these purposes owing to its high sensitivity to the intensity variations. For the analysis of U-burst key properties three reference points have been chosen. Point C indicates the turn over point of the burst while points A and B are placed on the ascending and descending legs of the burst at equal frequency shifts from the turning frequency (Figure 2b).

The frequency of points A and B should be as far from the turning frequency as possible allowing at the same time reliable determination of drift rates and times. In our opinion frequency of 70 MHz suits these requirements the best. At this frequency flux peaked at 9:34:32.0 UT (point A of ascending branch) and at 9:34:40.2 UT (point B of the descending one). Assuming plasma emission mechanism we may conclude that it takes 8.2 s for a beam to travel along the loop from point A to point B. The burst apex time (point C location on the time axis) is the time when drift rate of the burst equals 0. This time appeared to be 9:34:36 UT. Detection of correct location of point C on the frequency axis is a bit more complex due to not regular shape of the U burst spectrum at the turn over time (Figure 3). Thus taking the turn over frequency as the frequency of maximum intensity doesn't seem to be correct. Instead we took the median frequency of the spectral band bounded by the frequencies where the burst intensity falls by the background level. Taking into account the flatness of the top of the spectrum in Figure 3 we take the turnover frequency as 58±1 MHz. The absolute values of drift rates of the ascending and descending branches at the reference points A and B appeared to be equal to 4±0.2 MHz/s and just of opposite sign. This drift rate is four times of magnitude less than the drift rate of the preceding type III burst at the same frequency.

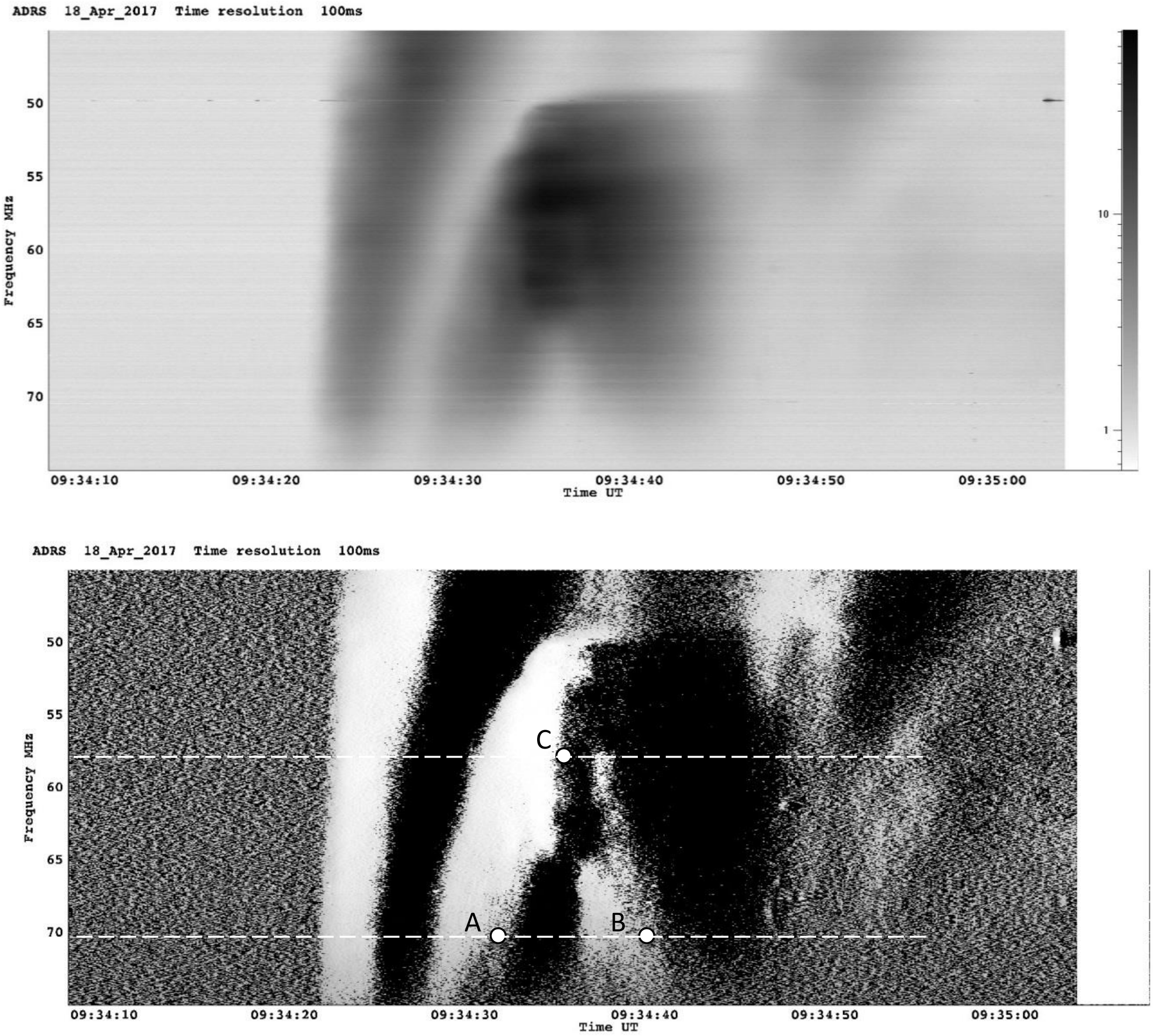

Figure 2. Inverted type U burst with equally developed legs: power dynamic spectrum (a) and differential in time dynamic spectrum (b). Points A and B are placed at the peaks of the ascending and descending legs of the burst at 70 MHz while point C denotes the burst apex.

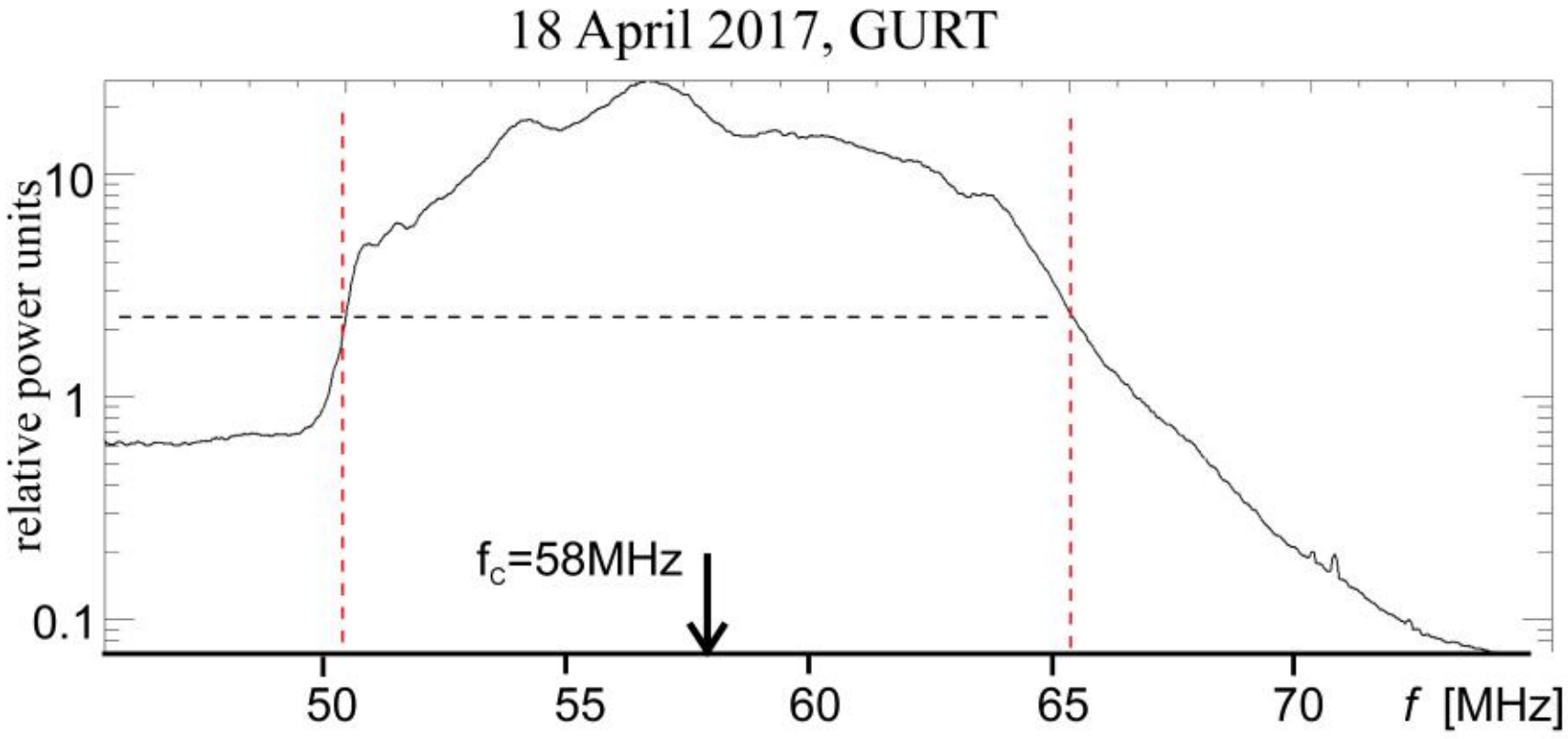

Figure 3. Inverted U-burst spectrum at 9:34:36 UT. Here vertical dashed lines denote the frequencies where the intensity falls by an order of magnitude with respect to the maximum.

**Analysis and discussion**

For given electron beam velocity the linear length of the top part of the coronal loop corresponding to the U burst segment bounded by the points A and B can be easily found. In type III bursts analysis the speed of the exciter is usually defined from the drift rate of the burst at given frequency supposing one of the known corona models. When heliographic data are absent the same corona models are also used for estimation of the heliocentric heights of the emission sources at different frequencies. In our opinion this method cannot be applied to U bursts because the confined inside the loop plasma is effectively isolated from the ambient plasma by the magnetic field of the loop. And according to (Reale, 2014) the temperature and density of the inside-the-loop plasmas may considerably differ from those outside the loop. Thus density profiles inside the loop in general may not follow any of commonly accepted models and we cannot estimate the exciter speed from the drift rate of the U burst.

At the same time Aurass and Klein (1997) showed that if the ascending branch of U burst and type III burst are observed within a few seconds from each other they most probably have common electron ejection source. In our case the ascending leg of the discussed U burst is preceded by the type III burst by 7.5 s. This type III burst with decay has been deeply analyzed by Melnik et al. (2019). The authors found that velocity of the electron beam responsible for the high frequency part of the type III burst (before the decay) was 0.2c if the fundamental emission was assumed and 0.33c for the harmonic one (c is the speed of light). There were several pros and cons to both the fundamental and harmonic emission mode. Nevertheless taking into account weak polarization and large solar longitude of the source the authors were inclined to the latter. This conclusion and the temporal proximity of the type III burst and the discussed type U bursts gives good reason to take the speed of electrons responsible for the type U generation equal to 0.33c.

The most plausible geometrical approximation of the upper part of the loop is a semi-ellipse lying in the plane parallel to the density gradient (Figure 4).

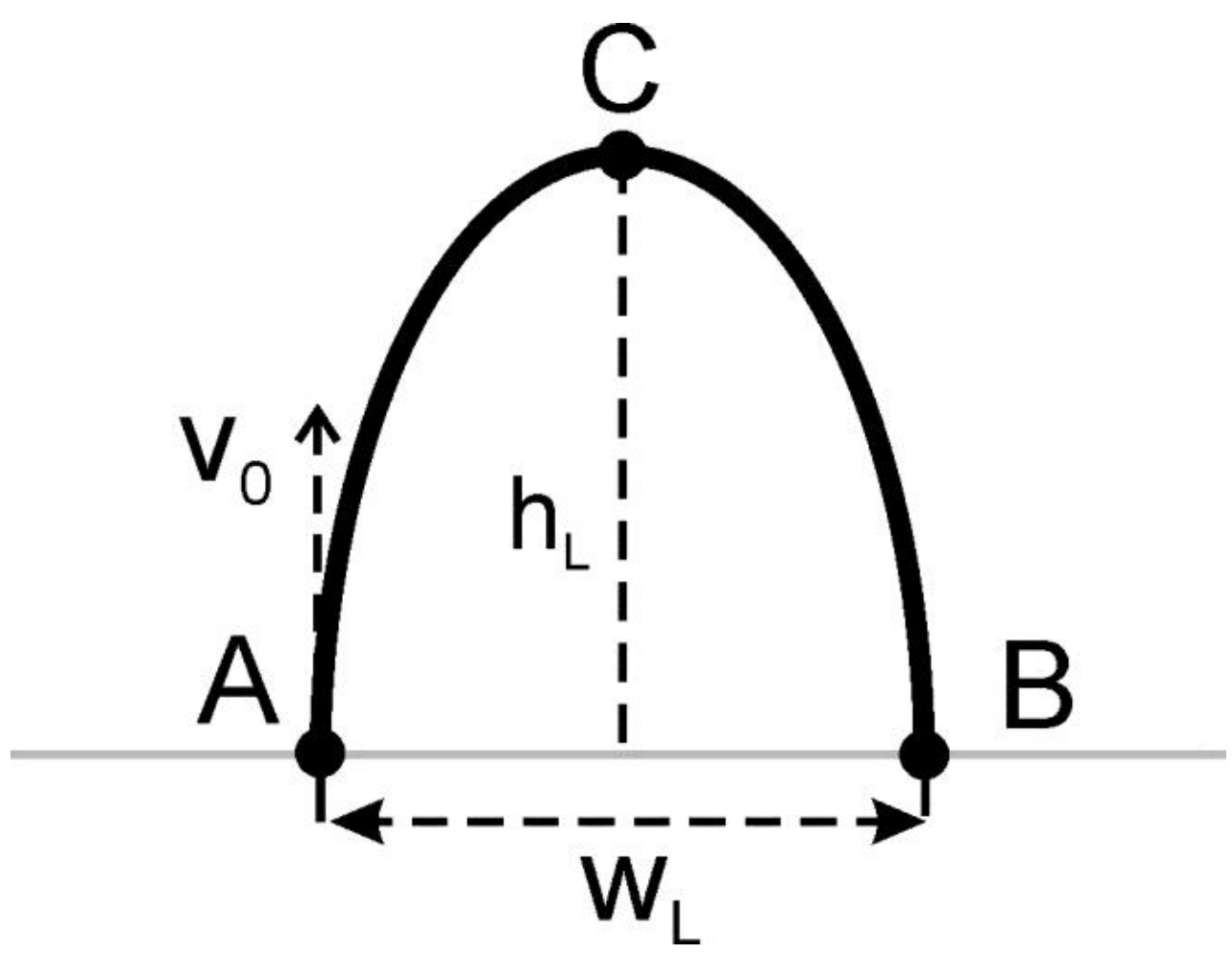

Figure 4. Geometric sketch of the loop segment corresponding to the observed U burst

Here points A, B and C correspond to the reference points of the same name on the U burst dynamic spectrum (see Figure 2b), $v_0$ is the electron beam velocity vector at point A, $h_L$ and $w_L$ denote the height and the width of the loop segment. Taking the electron velocity equal to 0.33c we define the linear length of the coronal loop segment $\boldsymbol{L}$ as 1.18 $R_\odot$, where $R_\odot$ is the solar radius. The values of $h_L$ and $w_L$ will be found from the further analysis.

It is commonly accepted that Type III electrons do not change their velocity while travelling along magnetic field lines (see for example Melnik et al., 1999; Melnik, et al., 2019). Though recent analysis (Krupar et al, 2015; Reid and Kontar, 2018) shows that at large distances from the Sun type III electrons use to decelerate with rate of about $3\cdot10^{-5}\boldsymbol{c}$ $cm^{-1}$, where c is the speed of light. Apparently for distances near 1 $R_\odot$ such deceleration can be neglected. Almost equal time intervals AC and CB and equal absolute drift rates of the observed U burst at points A and B in Figure 1b confirms this knowledge for electron beams moving along the top of the coronal loop.

The coronal loop is a closed magnetic structure which effectively isolates the confined inside it plasma from the ambient corona. Thus we may expect a considerable difference between the plasma parameters, namely temperature and density, inside and outside a loop (Reale, 2014).

Let's assume that the confined plasma is isothermal with density following Boltzmann relation in the form of as it was done in (Mann, 1999)

$$n(r) = n_0 exp\left(\frac{A}{r}\right) \tag{1}$$

where $n_0$ is the reference density, r is the distance from the solar center and A is the characteristic scale defined as

$$A = \frac{\tilde{\mu} G m_p M_\odot}{kT}. \tag{2}$$

Here $\tilde{\mu}$ – is the mean molecular weight, which equals 0.6 for solar corona (Priest, 1982), $G$ – is the gravitational constant, $m_p$ is the proton mass, $M_\odot$ is the mass of the Sun, $k$ is the Boltzmann constant and T is the temperature.

To determine the density profile for the loop of interest the values of T and $n_0$ are needed. As noted in (Reale, 2014) measurements of temperature is not a trivial task due to small optical thickness of coronal plasma and absence of direct measurements. Thus there is currently no unified view of the problem of the temperature distribution along a loop. Assuming that the plasma of the loop of interest is heated at its footpoints only it seems reasonable to take the isothermal plasma model as was suggested by Aschwanden (2001).

So we performed calculations for two fixed temperatures: 1.4 MK as it was taken by Mann et al. (1999) and twice as high temperature in order to find the basic relations between type U parameters and the properties of the associated coronal loop and the trends of their dependences on temperature. For given temperatures the characteristic scales A equal $6.92\cdot10^{11}$cm and $3.46\cdot10^{11}$cm respectively.

The value of $n_0$ can be retrieved from the drift rate of the burst as follows.

In general for plasma emission mechanism the drift rate of the burst can be defined as

$$\dot{f} = \frac{df}{dt} = \frac{df}{dn} \cdot \frac{dn}{dr} \cdot \frac{dr}{dt}, \tag{3}$$

Assuming the harmonic emission mode the first term of the equation (3) is

$$\frac{df}{dn} = \frac{f_p}{n_e}, \tag{4}$$

where $f_p$ is the local plasma frequency at the source of radio emission, $m_e$ is the electron mass, $e$ is the electron charge and $n_e$ is the electron density.

The second term is retrieved from (1) as

$$\frac{dn}{dr} = -n_e \cdot \frac{A}{r^2}. \tag{5}$$

And the third term is the radial component of the electron beam velocity $v_r = v \cdot cos\theta$, where $\theta$ is the angle between the velocity vector and density gradient.

Finally taking into account equations (4, 5) we can rewrite the equation (3) as

$$|\dot{f}| = \frac{A \cdot f_p \cdot v_0 \cos\theta}{r^2}. \tag{6}$$

In equation (6) and hereinafter we will operate with absolute values of the frequency drift rate since sign doesn't play role in further calculations.

After rewriting Equation (6) for ***r*** we obtain

$$r = \sqrt{\frac{A \cdot f_p \cdot v_0 \cdot \cos\theta}{\dot{f}}}. \tag{7}$$

Equation (7) allows us to find the height of the radio emission source using measured drift rate of the U burst at given frequency. As was said earlier we have chosen the reference points A and B at frequency 70MHz on both legs of the U burst. For the assumed geometrical approximation of the top of the loop angle $\theta$ =0. Taking $v_0$=0.33c, T=1.4 MK, $\dot{f}$=4 MHz/s $f_p$=35 MHz and assuming harmonic emission mode we obtain the heliocentric height of points A and B in the loop $r_A = r_B = 3.55 R_\odot$ .

Then knowing the exact height of the sources A and B of radio emission and the local plasma density at the source it is possible to find the reference density $n_0$ in equation (1). It equals $9 \cdot 10^5$ cm$^{-3}$. Thus we get the density dependence on heliocentric heights for the individual coronal loop associated with the observed type U burst in the form

$$n(r) = 9 \cdot 10^5 exp\left(\frac{10}{R}\right) \text{ cm}^{-3}, \tag{8}$$

where $R$ is the heliocentric height expressed in solar radii.

The height of points A and B in Figure 4 has been already determined thus we need to find the height of point C to complete the geometry of the loop segment. It can be done using Equation 8 which determines the local plasma frequency at every point inside the loop. Assuming harmonic emission mode the plasma frequency at point C should equal 29 MHz. From Equation 8 it follows that the heliocentric height of the loop apex corresponding to the U burst turnover frequency is $4.1R_{\odot}$. Thus the elevation of the point C above points A and B, i.e. the value of $h_L$ in Figure 4 equals $0.55R_{\odot}$. For a given half-perimeter of the ellipse and one of its semi-axes the width of the loop $w_L$ (Figure 4) was found to be 0.26 $R_{\odot}$.The appearance of the top of the loop for temperature T=1.4 MK and normal orientation of the loop is shown in Figure 5 with red color. In this case plasma inside the loop at points A and B is about 20 times as high as the surrounding plasma which is supposed to follow the Saito equation (Saito et al., 1970). Very close values were obtained for plasma enhancements inside much lower loops observed in X-rays (Stewart and Vorpahl, 1977). The authors found that the plasma inside a loop being $0.1R_{\odot}$ high was 16 times denser than outside it.

The calculations for the twice as high temperature T=2.8 MK give the value $n_0 = 2.06 \cdot 10^6$ cm$^{-3}$ and heliocentric height of the points A and B as 2.51 $R_{\odot}$. The loop apex in this case reaches the height of ~3.09 $R_{\odot}$ from the solar center as shown in Figure 5 with yellow curve. Then the height and the width of the top part of the loop (values $h_L$ and $w_L$ in Figure 4) are ~0.58 $R_{\odot}$ and ~0.11 $R_{\odot}$, respectively. It means that the apex of hotter loop is located lower in the corona. The loop segment connecting points A, C, and B itself is slightly higher and narrower than that for the cooler loop. Plasma density in the hotter loop is only ~7 times denser than that expected from the Saito's model at the same height.

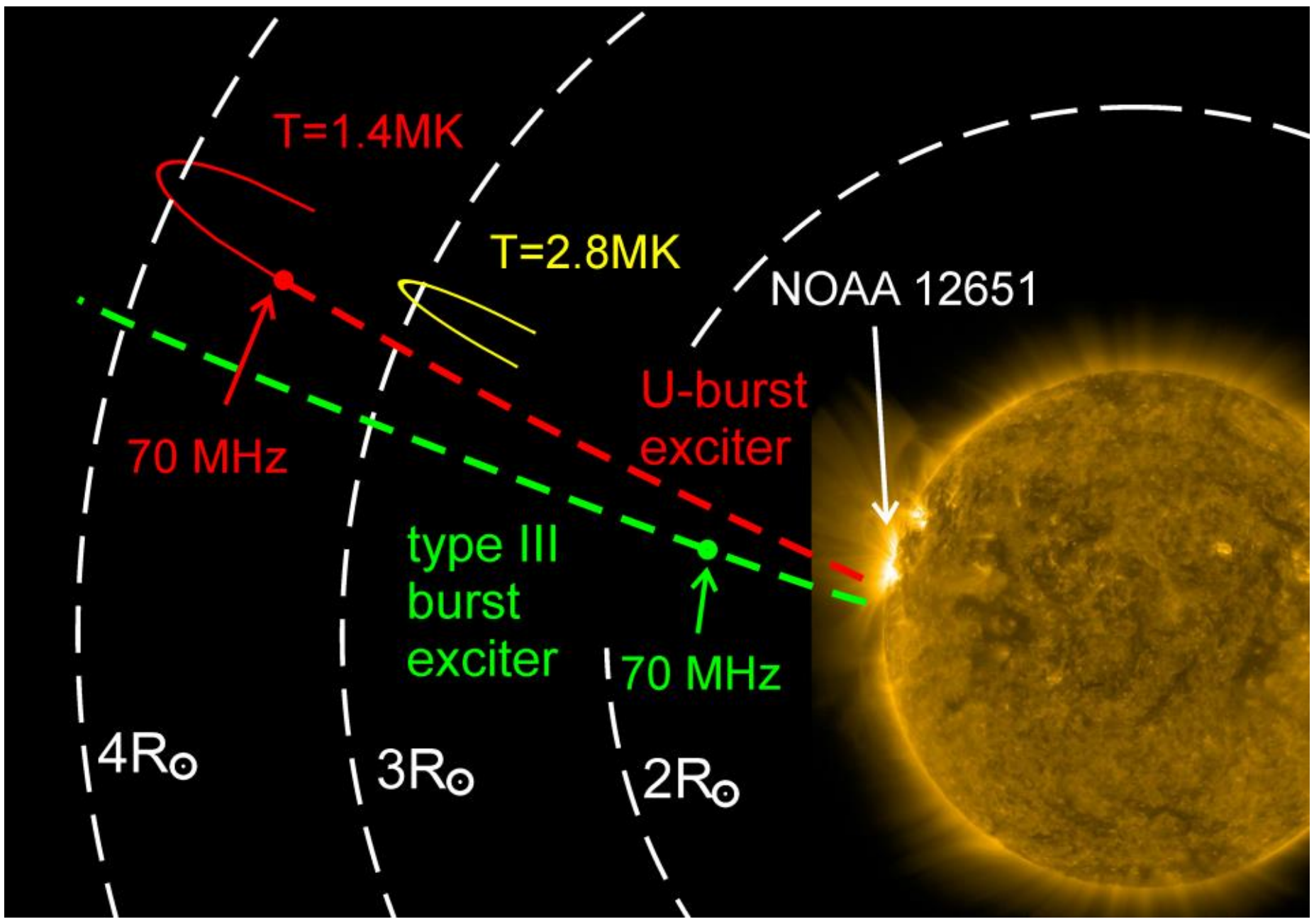

Figure 5. Trajectories of the type III electrons (green line) and inverted U-burst electrons (red line for plasma temperature 1.4MK and yellow line for 2.8MK). Solid curves correspond to the observed type U burst. The locations of type III and type U bursts sources at frequency of 70 MHz are denoted with solid circles.

The loop approximation by the normally directed semi-ellipse is of course a special case. In general coronal loop may deviate from the gradient direction. The analysis shows that tilted semi-ellipse may also fit the parameters of the observed type U bursts in such a way that heliocentric heights of all points in the loop are decreasing with increasing declination angle. For example if the loop with temperature of 1.4 MK is tilted by 45° to the radial direction its top will be as high as 3.36 $R_{\odot}$, what is about 18% less than in the case of normal loop. In this case the difference between the plasma densities inside and outside the loop also decreases. In addition the inclination of the loop plane results in slight increase of the semi-ellipse aspect ratio. For the above case the ellipticity constant increase from 0.24 for normal loop to 0.31 for the inclined one.

Finally let's suppose that the inverted type U burst and preceding type III burst have been generated by the same electron beam with velocity of 0.33c. The case when the electron beam could be injected simultaneously into one or more closed and open magnetic field structures was supposed earlier (Caroubalos, Couturier, and Prokaxis, 1973; Dorovskyy et al., 2010). According to our model for normally oriented loop containing 1.4 MK plasma and the harmonic emission mode the heliocentric height of the source of radio emission of the U-burst at 70 MHz equals 3.55 $R_{\odot}$. The source of preceding type III burst emission at the same frequency is located at 1.5 $R_{\odot}$ assuming harmonic emission and the Saito corona model (Saito et al., 1970). Thus it will take ~4.7 s for electromagnetic wave to travel from type III source to the height of 3.55$R_{\odot}$ where the U burst source of the same frequency is located. At the same time the exciter moving with the velocity of 0.33c will cover the distance between two sources for ~14.1 s. The total delay of the U burst maximum with respect to type III maximum at frequency of 70 MHz appears to be ~9.6 s and is very close to the observed delay of 7.5 s. Taking into account that all heights obtained in the present analysis are model dependent the hypothesis of common exciter for both of these bursts seems plausible. Apparently if this assumption is true then the delay of the U burst with respect to the preceding Type III burst together with spectral parameters of these bursts unambiguously define the temperature and the height of the loop. In our particular case the observed delay corresponds to the temperature of the inside-the-loop plasma of 1.9 MK. Consequently the top of this loop appears to be 3.6 $R_{\odot}$.

It is evident that temporal and spectral parameters of individual U burst with equally developed legs cannot provide exact values characterizing the physical properties and the geometry of the associated coronal loop. At the same time detailed analysis of the dynamic spectra of such bursts give us general relation between the height of the loop and the temperature of the confined plasma in such a way that lower loops contain hotter plasma and vice versa. This conclusion agrees well with the results received in (Dorovskyy, et al., 2015) from analysis of the U burst harmonic pair mutual delay. Thus knowing one of these two parameters we can define another one. From the current state of available observation facilities, such as interferometric observations with UTR-2 radio telescope (Dorovskyy, et al., 2018; Melnik, et al., 2018) or tied-array imaging realized at LOFAR station (Reid and Kontar, 2017; Gordovskyy et al., 2019) it seems more realistic to define the height of the loop as well as its deviation from the normal direction. This will enable us to yield the temperature of the confined plasma, which is difficult to measure by other means at such heights.

The proposed model also explains systematically lower drift rates of the ascending branches of type U bursts compared to the drift rates of standard type III bursts at the same frequency. The sources of U bursts are located much higher in the corona than the sources of type III bursts at the same frequency, and then density gradient at the U burst source is less resulting in lower drift rates.

## Conclusions

The unique inverted type U burst with equally developed ascending and descending legs was observed on 18 April 2017 by one section of radio telescope GURT. This allowed to obtain the extended information about the coronal loop from the parameters of the associated type U burst even with lack of radio source imaging. The loop apparently originated from the limb active region NOAA 12651. Under the assumption of a planar loop of semi-elliptic shape and gravitational stratification of the isothermal plasma inside the loop we found that the heliocentric height of the source of radio emission is unambiguously defined by the electromagnetic wave frequency, the frequency drift rate of the U burst at corresponding frequency, the plasma temperature and the inclination of the loop plane from the radial direction. For the case of the observed U-burst, radial orientation of the loop plane and temperature equal to 1.4 MK the heliocentric height of the top of the loop was found to be 4.1 $R_\odot$. From this result it follows that the sources of U bursts are in general higher than those of type III bursts at the same frequency. Equally developed legs of the burst enable to find the length of the loop segment corresponding to the visible type U burst as 1.18 $R_\odot$. It was also shown that the electron beam responsible for this type U burst generation did not accelerate or decelerate while travelling along the loop. In our opinion this electron beam is also responsible for the preceding type III burst with decay. In addition we found that with other things being equal the lower loops contain hotter plasmas and vice versa, that agrees well with previously obtained results for U bursts harmonic pair (Dorovskyy, et al., 2015). It is clear from the analysis that determination of one of mutually connected parameters, e.g. temperature or height of the loop by some other independent method will unambiguously reveal the other one. Thus complementation of traditional high resolution spectral analysis by retrieving of spatial parameters of the U bursts from interferometric observations or tied-array imaging is highly desirable.

The analysis shows that the inverted type U bursts and preceding by 7.5 s type III burst could be generated by the same fast electron beam injected simultaneously along open and closed magnetic structures.

One more important result of the obtained in the paper is the evidence that the electron beams preserve their speed while travelling along high coronal loops at least within distances around 1 $R_\odot$ and even moving towards the Sun.

Our model also suggests that U bursts with lower turning frequencies, e.g. around 25 MHz may originate from the coronal loops which are ~13$R_\odot$ high. Such loops can be the subject of both remote observations by ground-based highly efficient radio telescopes (e.g. LOFAR, NenuFAR, UTR-2 etc) and in-situ analysis by Parker Solar Probe (PSP) during its closest perihelia.

**Acknowledgments**. The research was supported by projects Radiotelescope (0120U100234) and Spectr-3 (0117U000245) of the National Academy of Sciences of Ukraine.